\documentclass[doublecol]{epl2}

\title{The Effect of Pinning on Drag in Coupled One-Dimensional Channels of Particles} 
\shorttitle{Effect of Pinning on Drag in Coupled One-Dimensional Particle Systems}
\author{C. Bairnsfather, C.J. Olson Reichhardt and C. Reichhardt} 
\shortauthor{C. Bairnsfather, C.J.O. Reichhardt and C. Reichhardt}

\institute{ 
Theoretical Division, 
Los Alamos National Laboratory, Los Alamos, New Mexico 87545\\
Department of Physics, Purdue University, West Lafayette, Indiana 47907
}

\abstract{
We consider a simple model
for examining the effects of quenched disorder on drag consisting of
particles interacting via a Yukawa potential that are placed in
two coupled one-dimensional channels.
The particles in one channel
are driven and experience a drag from the undriven particles in the second
channel.
In the absence of pinning, for a finite driving force
there is no pinned phase; instead, there are two dynamical regimes of 
completely 
coupled or locked flow and partially coupled flow. 
When pinning is added to one or both channels, we find
that a remarkably rich variety of dynamical phases and drag effects
arise that can be clearly identified by features in the
velocity force curves.  
The presence of quenched disorder 
in only the undriven channel can induce a pinned 
phase in both channels.  Above the depinning transition, 
the drag on the driven particles decreases with 
increasing pinning strength, and for high enough 
pinning strength, the particles in the undriven channel reach 
a reentrant pinned phase which produces 
a complete decoupling of the channels.    
We map out the dynamic phase diagrams 
as a function of pinning strength and the density of pinning in each channel.
Our results may be relevant for understanding 
drag coupling in 1D Wigner crystal phases, and 
the effects we observe could also be explored 
using colloids in coupled channels produced
with optical arrays, 
vortices in nanostructured superconductors,
or other layered systems where drag effects arise.  
}

\begin{document}

\maketitle
There are many examples 
of one- and two-dimensional (1D and 2D)
coupled bilayers or coupled channels of interacting particles,
including vortices in superconducting bilayers \cite{G,Clem},
colloidal systems \cite{Colloid,Colloid2,Colloid3,Colloid4}, 
dusty plasmas \cite{Donko}, 
and Wigner crystals \cite{R,Stopa,B}.  
In many of these systems it is possible to apply 
an external drive to one of the layers and measure the 
resulting response of the other layer 
as well as the drag effect produced by the particles in the undriven
layer on those in the driven layer.
For instance, this type of measurement has been performed for
the transformer geometry 
in two-layer superconducting vortex systems \cite{G,Clem}. 
If the vortices in each layer  
are completely coupled, the measured 
response is the same in both layers.  If instead the vortices are
only partially coupled  between layers, the response is reduced
in the undriven layer compared to the driven layer.
A similar coupling-decoupling transition is also predicted
for coupled wires containing 1D Wigner crystals \cite{R}, and 
certain drag effects in 1D wires have 
been interpreted as resulting from the
formation of 1D Wigner crystal states \cite{Stopa}.               

Many of these systems can contain some form of quenched disorder which 
could produce pinning effects \cite{Peeters2}; however, very little
is known about how pinning alters 
the drag or transport properties in layered geometries.  
The quenched disorder could be strong in one channel
and weak in another, or it could be of equal strength in both channels, 
resulting in different types of dynamic phases.
Particles driven over random quenched
disorder in single layer systems
are known to exhibit a variety of 
dynamical phases associated with distinct transport signatures, as shown
in simulations \cite{Peeters2,Olson,Reichhardt,Noise} and experiments \cite{S}. 
These studies suggest that
distinct dynamics could also arise 
in a two-layer drag system when pinning is present.
We note that there have been studies 
of two-layer driven colloidal systems; however, the effect of
quenched disorder was not considered \cite{Colloid2}.
  
In this work we propose a simple
model of two coupled 1D channels of particles 
with repulsive Yukawa interactions
where there a drive channel and a drag channel, 
and where pinning is added to one or both channels.
Despite the apparent simplicity of this system, 
we find that a remarkably rich variety of distinct dynamical phases 
are possible which produce pronounced changes in the 
transport and drag effects. 
Although our model treats 1D channels, we expect that many of the same 
effects should be generic to coupled layer systems with 
quenched disorder. 
The specific system we consider could be realized experimentally using
colloidal particles
in channel geometries \cite{Koppl} 
where a driving force is applied to one channel via optical means or
an electric field. 
The effects we observe could also be studied with
coupled 1D Wigner crystals \cite{Peeters1,Peeters2,Stopa,Bockrath} or
in superconductors with nanofabricated channel geometries \cite{Kes} when 
the current is applied to only one channel 
and the response is measured in both channels.    
In our system, we find that without pinning
there is a 
finite drive transition from a locked regime to a partially decoupled 
regime in which the response of the drag channel decreases with
increasing drive.
Addition of pinning to the drag channel
can induce a pinned phase for both channels.
As the drive increases, both channels depin into a locked or partially locked
phase, followed at high drives
by a reentrant pinning of the drag channel when the
dynamic coupling of the two channels becomes weak enough.
When the driving force is fixed at a low value, increasing the strength of
the pinning in the drag channel increases the effective drag on the particles
in the driven channel; for higher fixed driving force, however, increasing
the pinning strength reduces the drag on the driven channel and causes the
channels to decouple when the drag channel becomes reentrantly pinned.
We also observe strong fluctuations in both channels 
at the locked to partially locked transition
for intermediate strengths of quenched disorder. 
We map out the evolution of the dynamical phases 
for a wide variety of parameters including 
pinning in both channels, pinning in only one of the channels,
and for varied pinning densities. 

\begin{figure}
\includegraphics[width=\columnwidth]{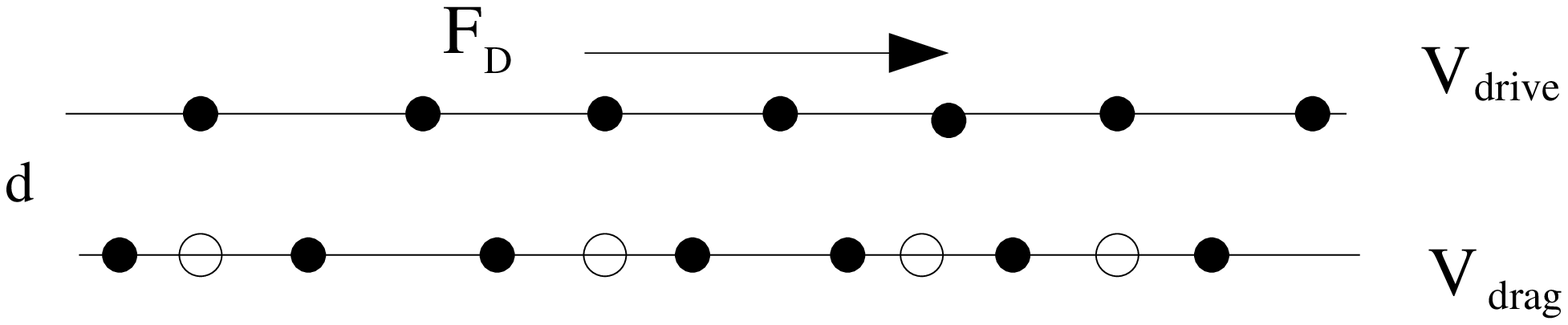}
\caption{
A schematic of the system of two coupled 1D channels separated by a distance 
$d$. Each channel contains $N$ particles (black dots) which interact via
a Yukawa potential with particles in the same and the adjacent channel.
An external drive $F_D$ is applied to the particles in the 
drive channel, resulting in an average particle 
velocity of $V_{drive}$ in the drive channel. 
The velocity response  in the drag channel is $V_{drag}$. 
Quenched disorder is introduced in the form of 
localized pinning sites (open circles) 
of maximum force $F_{p}$ which are spaced randomly in the channel without
overlapping. 
The pinning can be placed only in the drag channel (as shown),
only in the drive channel, or in both channels.
}
\label{fig_schematics}
\end{figure}

{\it Simulation-} In fig.~1 we show a schematic of our system 
which consists of two 1D channels 
separated by a distance $d$ 
with periodic boundary conditions in the $x$-direction.
Each channel contains $N$ particles which interact via a repulsive Yukawa
potential with other particles in the same channel and with particles in
adjacent channels.
Particles in the upper or drive channel 
are subjected to an applied external force $F_{D}$.   
A single particle $i$ located at position ${\bf R}_i$ 
undergoes motion obtained by integrating the  overdamped equation of motion:
\begin{equation}
\eta\frac{ d{\bf R}_{i}}{dt} = {\bf F}_{i}^{pp}  
+ {\bf F}^{s}_{i} + {\bf F}^{D}_{i} .
\end{equation}
Here $\eta$ is the damping constant which is set to 
$\eta=1.0$. The particle-particle interaction force is
${\bf F}^{pp} = \sum^{2N}_{j\ne i}-\nabla V(R_{ij})$ 
where
$V(R_{ij}) = (E_{0}/R_{ij})\exp(-\kappa R_{ij}){\bf \hat R}_{ij}$,
$R_{ij}=|{\bf R}_i-{\bf R}_j|$, ${\bf \hat R}_{ij}=({\bf R}_i-{\bf R}_j)/R_{ij}$,
and $E_{0} = Z^{*2}/4\pi\epsilon\epsilon_{0}a_{0}$.  
Here $\epsilon$ is the solvent dielectric constant, 
$Z^{*}$ is the effective charge,
and $1/\kappa$ is the screening length. 
In each channel, the average particle spacing is $a$ and we take $d/a=2/3$.
We also take $1/\kappa=2d$ to ensure that coupling between the two channels
is possible.
We note that we have also considered other values of $d$, $a$, and 
$\kappa$ and find the same qualitative features, 
indicating that our results should be generic for this class
of system.  
The pinning force ${\bf F}_i^{s}$ arises from $N_p$ 
attractive parabolic potentials
with a maximum force of $F_p$ and a radius of $R_p=0.5$,
${\bf F}_i^{p}=\sum_{k=1}^{N_p}E_0F_p(R_{ik}^{p}/R_p)\Theta(R_p-R_{ik})
{\bf {\hat R}}_{ik}^p$.
Here $\Theta$ is the Heaviside step function, ${\bf R}_k^p$ is the location of
pinning site $k$, $R_{ik}^p=|{\bf R}_i-{\bf R}_k^p|$, and
${\bf {\hat R}}_{ik}^p=({\bf R}_i-{\bf R}_k^p)/R_{ik}^p$.
In this work we consider 
the limit in which the number of pinning sites is smaller than 
or similar to the number of
particles in the system; however, 
we find the same qualitative features 
when we change the pinning density or strength.  
The external force $F_{D}$ is applied only to the particles 
in the driven channel.  
We increase $F_D$ from zero in small increments of 
$\delta F_{D} = 0.001$, spending a waiting time 
of $10^4$ simulation time steps at each increment.
We measure the average velocity of the particles in the drive and drag channels,
$V_{drag}=N_{drag}^{-1}\sum_{i=1}^{N_{drag}}d{\bf \hat x}_i/dt$ and
$V_{drive}=N_{drive}^{-1}\sum_{i=1}^{N_{drive}}d{\bf \hat x}_i/dt$,
where $N_{drag}=N$ is the number of particles in the drag channel and 
$N_{drive}=N$ is the number of particles in the drive channel.
This simulation method, employing overdamped dynamics with random pinning,
has been used previously to explore the behavior
of colloids driven over random and periodic substrates 
as well as two layer systems.   
A similar method has been used to study
sliding Wigner crystals and vortices in type-II superconductors.

\begin{figure}
\includegraphics[width=\columnwidth]{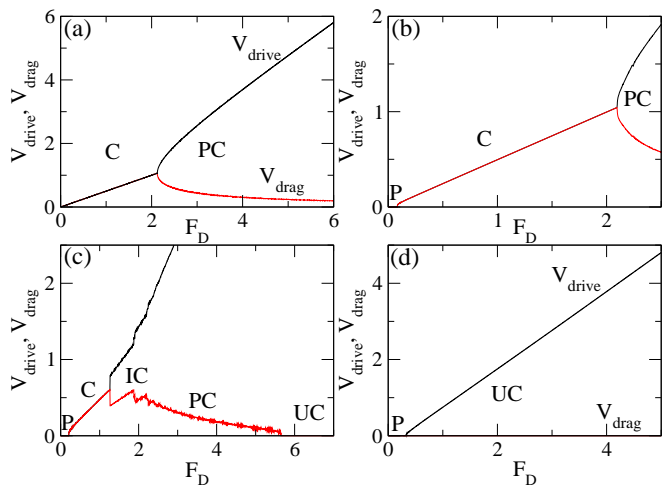}
\caption{The average particle velocity in the drive channel 
$V_{drive}$ and the drag channel $V_{drag}$ vs applied force $F_{D}$ for
a sample with pinning in the drag channel at $N_p/N=0.083$.
(a) The pin-free case $F_{p} = 0.0$ has two regions: an initial
coupled region ($C$) where motion in both channels is completely locked, and 
a partially coupled ($PC$) region 
where $V_{drag}$ monotonically decreases while $V_{drive}$ increases with
increasing $F_D$.
(b) For $F_{p} = 1.0$
there is an initial pinned region ($P$) in both channels in addition to
the $C$ and $PC$ regions.
(c) At $F_{p} = 2.5$,
an intermittent region ($IC$) containing a series of asymmetric jumps in
$V_{drag}$ appears between the $C$ and the $PC$ regions.
At $F_D=5.66$
there is a reentrant pinning of the drag channel 
above which $V_{drag} = 0.0$, producing an uncoupled region ($UC$).
(d) At $F_{p} =4.0$, only the $P$ and $UC$ regions appear
and $V_{drag}=0$ for all $F_D$.
}
\label{fig_grainboundaries}
\end{figure}

In fig.~2(a) we show the velocity force curves 
$V_{drive}$ and $V_{drag}$ versus $F_D$ for the 
case where there is no pinning in either channel.
At low $F_D$, $V_{drive}=V_{drag}$, indicating that
the particles are moving together in the coupled region ($C$). 
At $F_{D} = 2.0$, 
the average particle velocity 
in each channel is just under $1.0$. In the absence
of the drag channel, we would have $V_{drive}=2.0$ at $F_{D}=2.0$
since $V = F_{D}/\eta$ under our overdamped dynamics. 
When the drive channel is locked with the drag channel,
the same driving force must be used to transport twice as many particles,
reducing the particle velocity by half.
At
$F_{D} = 2.124$ there is a transition to a partially 
coupled ($PC$) state where
the velocity curves branch apart. 
$V_{drive}$ continues to increase with increasing $F_D$ with a slope
higher than the slope of $V_{drive}$ below $F_D=2.124$, while
$V_{drag}$ decreases with increasing $F_D$. 
Since there is no pinning, 
in the $PC$ region $V_{drag}$ gets smaller and smaller with increasing
$F_D$ but never reaches zero.
The general shape of the velocity force curves
is very similar to the response observed for the 
superconducting transformer geometry where vortices in a pair of adjacent
superconducting layers can couple and decouple depending on the
applied current \cite{G,Clem}.  In the superconducting system,
the voltage response is proportional to the particle velocity and the applied
current is proportional to $F_{D}$.
Additionally, the velocity-force curves
in fig.~2(a) are in agreement with the predicted 
coupling-decoupling transition for 
disorder-free coupled 1D Wigner crystal systems \cite{R}.      

In fig.~2(b) we plot $V_{drive}$ and $V_{drag}$ versus $F_D$ for a system
containing pinning in the drag channel with $N_p/N=0.083$ and $F_p=1.0$.
The curves are similar to the pin free case, but
a pinned region ($P$) now appears 
for $F_{D} < 0.1$ when the quenched disorder in the drag channel
causes the particles in both channels to stop moving.
The particles in the drag channel are directly pinned by the pinning
sites, but the particles in the driven channel are only indirectly pinned by
their interactions with the particles in the drag channel, which
produce an effective periodic pinning potential for the driven particles.
The driven particles can either simply move over this periodic potential,
in which case we find a decoupled region ($UC$) where
the driven particles are moving and
the drag particles are pinned,
or the driven particles can drag the periodic
potential along with them, depinning the drag particles and
producing a moving coupled region.
As $F_D$ increases, there is a transition from the coupled to partially
coupled flow at the same $F_D=2.124$ as in the pin-free system.
Figure~2(c) shows the velocity-force curves for 
a sample with $F_{p} = 2.5$, where 
the transition out of region $C$ falls at a lower $F_{D} = 1.26$ 
and is followed by a region containing
a series of jumps in both $V_{drive}$ and $V_{drag}$.
In this intermittently coupled ($IC$) region,
each downward jump in $V_{drag}$ 
is preceded by a range of $F_D$ over which $V_{drag}$ increases
linearly with increasing $F_D$.  
The jumps in $V_{drag}$ are accompanied by similar sharp upward jumps
in the value of $V_{drive}$.
As $F_D$ increases, the jumps decrease in size until
the system crosses over to the partially coupled ($PC$) region where
$V_{drag}$ decreases linearly and smoothly with increasing $F_D$.
At high $F_D$, the uncoupled channel ($UC$) region 
appears when the particles in the drag channel
become pinned again for $F_D \geq 5.66$ but the particles in the driven
channel continue to flow.
The reentrant pinning of the drag channels results when the effective 
dynamical coupling between the two channels weakens for higher $F_D$, as
indicated by the decrease in $V_{drag}$ with
increasing $F_{D}$ in the $PC$ region of the pin-free sample
shown in fig.~2(a). 
When the dynamical coupling between the channels drops below a critical value,
the drag force experienced by the particles in the drag
channel falls below the depinning threshold for the pinning sites in 
the drag channel, resulting in the repinning transition.
We expect this reentrant pinning effect to
be a generic feature in drag systems containing quenched disorder, and its 
signature is a sudden decrease of the drag channel response to zero.
In fig.~2(d) for $F_{p} = 4.0$, the drag channel never depins over the
range of $F_D$ shown, and 
there is a single transition 
from region $P$ where both channels are pinned to the 
decoupled region $UC$.    

\begin{figure}
\includegraphics[width=\columnwidth]{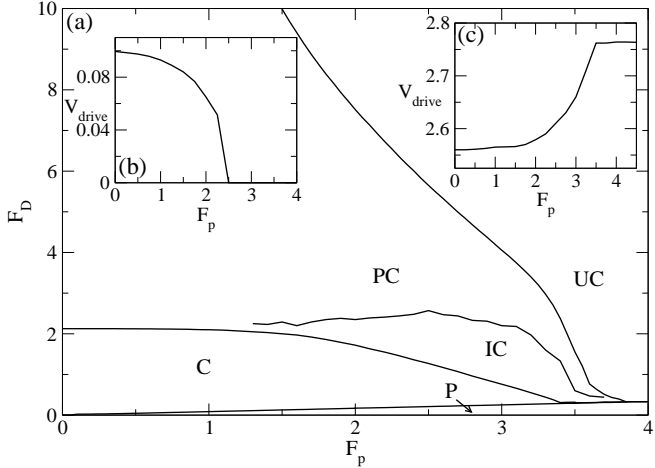}
\caption{
(a) The dynamic phase diagram of $F_{D}$ vs $F_{p}$ for the system in 
fig.~2 for pinning in only the drag channel.
Each region is marked: pinned ($P$), coupled ($C$), intermittently coupled
($IC$), partially coupled ($PC$), and uncoupled ($UC$).
(b) $V_{drive}$ versus $F_{p}$ from the system in (a) 
at $F_{D} = 0.2$.  The increasing
effective damping $\eta^*=F_D/V_{drive}$ 
produces a monotonic decrease in $V_{drive}$ with increasing
$F_p$ until the particles in the driven channel repin and the velocity
drops to zero in region $P$.
(c) $V_{drive}$ vs $F_{D}$ 
for the system in (a) at $F_{D} = 3.0$. 
Here $V_{drive}$ 
increases with increasing $F_{p}$ 
until the onset of the $UC$ phase, above which $V_{drive}$ saturates to
a constant value.
}
\label{fig_plots}
\end{figure}

In fig.~3(a) we plot the dynamic phase diagram 
$F_{D}$ versus $F_{p}$ for the system in fig.~2 containing pinning
only in the drag channel.
The transition between the $PC$ 
and $UC$ regions drops to lower values of $F_{D}$ 
with increasing $F_{p}$ since 
a lower drive is required to reduce the effective drag
force below the depinning threshold of the drag channel at higher $F_p$.
For $F_{p}>  3.88$,  only the $P$ and $UC$ regions appear. 
The extent of region $C$ decreases with increasing $F_p$ until region $C$
disappears
for $F_{p} > 3.45$. 
The $IC$ region vanishes
at higher values of $F_{D}$ and reaches its maximum extent
near $F_{p} = 2.95$. 
In fig.~3(b) we plot $V_{drive}$ versus $F_p$ for 
the same system at fixed $F_D=0.2$.
Here $V_{drive}$ decreases with increasing $F_{P}$ 
until the system enters the pinned region at $F_{p} = 2.0$ and $V_{drive}$
drops to zero.
The effective damping of the driven particles, 
$\eta^{*} = F_{D}/V_{drive}$, increases with increasing $F_{p}$. 
In fig.~3(c) we show $V_{drive}$ versus $F_p$ at fixed $F_{D} = 3.0$, 
where $V_{drive}$ monotonically increases 
until the system enters the $UC$ phase, and $V_{drive}$ saturates to
a constant value.
These results show that 
pinning in the drag channel can either increase or decrease the
effective drag on the driven channel 
depending on the strength of the pinning and the amplitude
of the dc drive. 

\begin{figure}
\includegraphics[width=\columnwidth]{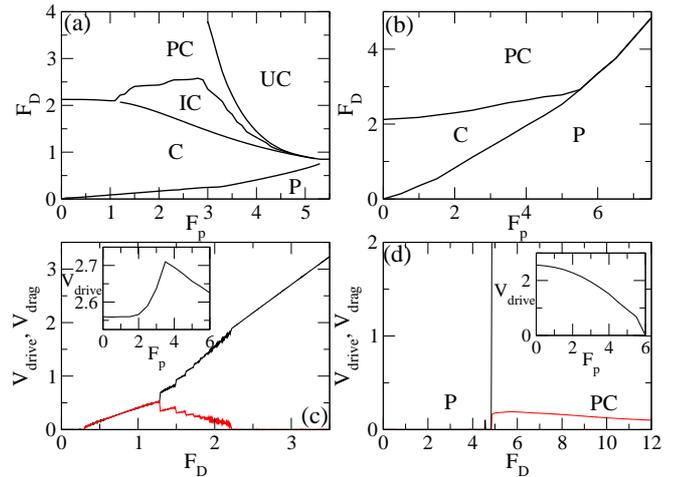}
\caption{
(a) The dynamic phase diagram 
of $F_{D}$ vs $F_{p}$ for a system with pinning in both channels.
Each channel has $N_p/N=0.083$.  The region labels are the same as in
fig.~3(a).
(b) The dynamic phase diagram of $F_D$ vs $F_p$ 
for a system with pinning in only the drive channel, 
with $N_p/N = 1.33$. Here the uncoupled channel region is
absent 
and there can be a direct transition from region $P$ to region $PC$.
(c) $V_{drive}$ (upper curve) and $V_{drag}$ (lower curve) 
vs $F_D$ for the system in (a) at $F_{p} = 4.0$.
Inset: $V_{drive}$ vs $F_p$ for the system in (a) at $F_D=3.0$.
(d) $V_{drive}$ (upper curve) and $V_{drag}$ (lower curve)
vs $F_D$ for the system in (b) at $F_{p} = 7.5$. 
Inset: $V_{drive}$ vs $F_{p}$ for the system in (b) at $F_{D} = 3.0$.
 }
\label{fig_four}
\end{figure}

We next consider a system containing pinning in both channels.
Figure 4(a) shows the dynamic phase diagram 
of $F_D$ versus $F_p$ for a sample with a pinning density of 
$N_p/N=0.083$ in each channel.
The overall shape of the phase diagram
resembles the phase diagram in fig.~3(a) 
obtained with pinning in only the drag channel. 
The pinned phase $P$ grows in size with increasing $F_p$ much more rapidly
when there is pinning in both channels than when there is pinning in only
the drag channel.
In fig.~4(c) we plot $V_{drive}$ and $V_{drag}$ versus $F_D$ at $F_p=4.0$
for the sample from fig.~4(a) with pinning in both channels.
When $V_{drag}$ drops to zero at $F_D=2.22$ at the transition to region $UC$,
there is a jump up in the value of $V_{drive}$. 
In fig.~4(b) we show the phase diagram 
of $F_D$ versus $F_P$ for a sample containing 
pinning in only the drive channel with
$N_p/N = 1.33$. 
The $UC$ phase no longer appears since the particles in the drag channel 
can only reentrantly repin when there is pinning in the drag channel.
For $F_{p} > 5.5$, the system crosses
directly from region $P$ to region $PC$ without passing through the 
$C$ region.
Figure 4(d) illustrates $V_{drive}$ and $V_{drag}$ versus $F_D$ for this
system at $F_p=7.5$.  There is a transition directly from region $P$
to region $PC$ at $F_D=4.8$, and above this drive $V_{drag}$ gradually
decreases with increasing $F_D$.

In the inset of fig.~4(c) we plot $V_{drive}$ versus 
$F_{p}$ at $F_{D} = 3.0$ for the system in fig.~4(a) where there
is pinning in both channels. 
For $F_p<3.5$, $V_{drag}$ increases with increasing $F_{p}$, just  
as shown in fig.~3(c) for the sample with pinning in only the drag channel.
For $F_p>3.5$, when there is pinning in both channels
$V_{drag}$ decreases with increasing $F_p$ for $F_p>3.5$ rather than
saturating.
The increase in $V_{drag}$ with increasing $F_p$ for $F_p<3.5$ results when
the pinning in the drag channel reduces the coupling between the driven and
drag particles.
For $F_{p} > 3.5$ at $F_{D} = 3.0$, the particles in the drag channel 
are always pinned.  If there is pinning in only the drag channel, 
$V_{drive}$ remains constant for $F_p>3.5$ as shown in fig.~3(c); however,
if pinning is also present in the drive channel, it produces an increasing
drag on the driven particles, causing $V_{drive}$ to decrease with increasing
$F_p$ at $F_p>3.5$ for the sample with pinning in both channels.
In the inset of fig.~4(d) we illustrate this effect more clearly by 
plotting $V_{drive}$ vs $F_{p}$ for a system
with pinning in only the drive channel. 
Here $V_{drive}$ monotonically decreases with increasing $F_p$ until it
reaches zero once all the drive particles become pinned.

\begin{figure}
\includegraphics[width=\columnwidth]{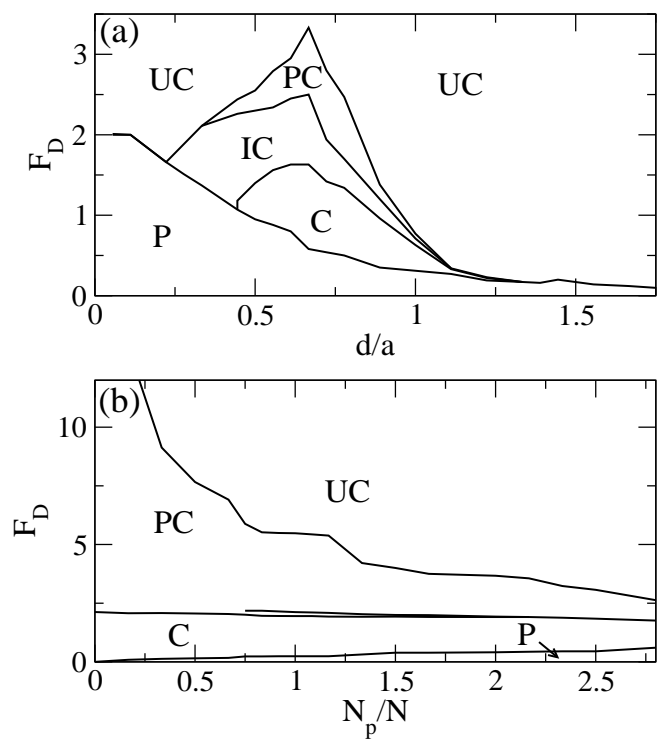}
\caption{
(a) The dynamic phase diagram of $F_{D}$ vs $d/a$ 
for a sample with pinning in both channels at $F_{p} = 2.0$
and with $N_{p}/N = 0.25$ at $d/a = 0.67$. 
(b) The dynamic phase diagram of $F_{D}$ vs $N_{p}/N$ for 
a sample with pinning in both channels
at $d/a = 0.67$ and $F_{p} = 1.0$.    
}
\label{fig_five}
\end{figure}

It is also possible to vary $d/a$, where $d$ is the
distance between the channels 
and $a$ is the average particle spacing within the channels. 
We consider the effect of varying particle density in 
samples with fixed $d$.
Increasing the particle density reduces the effective coupling between
the channels since the periodic potential experienced by the particles in
one channel produced by the particles in the other channel becomes smoother
as the particle density increases.
In fig.~5(a) we plot
the dynamic phase diagram for $F_{d}$ versus $d/a$ 
in a sample with pinning in both channels for fixed
$F_{p} = 2.0$ and
with $N_{p}/N = 0.25$ at $d/a = 0.67$.  
At low $d/a$, there is a large pinned region since the particles within
each channel are far from each other, strongly enhancing the 
effectiveness of an individual pinning site.
For $d/a<0.22$, the system passes directly from region $P$ to region $UC$.
For large $d/a$, 
each particle interacts strongly with other particles in the same channel
but the coupling to particles in the other channel is greatly reduced,
so the system again crosses directly from the pinned to the $UC$ region. 
For intermediate values of $d/a$, the three other regions $C$, $IC$, and
$PC$ appear in windows between the $P$ and $UC$ regions.  The total width of
these intermediate windows reaches a peak at $d/a=0.66$.
In fig.~5(b) we show the dynamic phase diagram of $F_D$ versus
the pinning density $N_p/N$ for a system with
pinning in only the drag channel at $F_{p} = 1.0$ and $d/a = 0.67$. 
The transition from region $PC$ to region $UC$ drops to lower $F_D$ with
increasing $N_p/N$ since higher pinning density causes the repinning 
threshold in the drag channel to decrease.
The transition between region $C$ and region $PC$ also shifts to
lower $F_D$ with increasing $N_p/N$,
and there is a narrow window (not labeled on the figure) 
where region $IC$ appears between regions $C$ and $PC$. 
There is also a small plateau in all of the transitions
near $N_{p}/N = 1.0$, which is a weak commensuration effect. 
For periodically arranged pinning
sites, we expect much stronger commensuration effects
to occur; these effects will be considered elsewhere.     

In summary, we introduce a simple 
model to study the effects of pinning on drag in coupled
one-dimensional channels of particles interacting via a Yukawa 
potential
where a drive is applied to only one channel. 
This system exhibits a rich variety of dynamical phases which are 
associated with distinct features in the velocity-force curves. 
In the absence of pinning, at low drives both channels
are coupled and exhibit identical particle velocities.  As the drive is
increased, there is a transition to
a partially coupled phase where the velocity
in the drag channel deceases with increasing drive 
while the velocity in the driven channel increases with
increasing drive.
Placing pinning in only the drag channel 
produces a region in which both channels are pinned
as well as a reentrant pinning of the particles in the drag channel 
at higher drives when the coupling between the channels is dynamically
weakened.
When strong pinning is placed either in the drag channel or in both
channels, we observe 
only a pinned region and a decoupled region.
For intermediate pinning strength,
we find a strongly fluctuating region 
where a series of asymmetric jumps appear in the 
velocity-force curves. 
Increasing the strength of the pinning in the drag channel 
generally reduces the effective drag on the particles in 
the driven channel at higher drives since the pinned particles in the
drag channel are unable to absorb momentum 
from the particles in the drive channel. 
We map the dynamic phase diagrams for samples with
pinning in both channels, pinning in only the drive channel, and pinning
in only the drag channel.  The smallest number of phases appears for
pinning in only the drive channel, while samples with pinning in both 
channels exhibit multiple dynamical phase transitions even when the
particle density or pinning density is altered, showing that the effects 
we observe are robust for a wide range of parameters.   
Our results could be important for understanding 
drag effects in systems where 1D Wigner crystal states
can occur and for understanding the general effects of quenched disorder 
in other drag and transformer geometries.  
Additionally, our system could be directly realized experimentally 
for colloidal or dusty plasma systems in coupled 1D channels. 
Extensions of this model that could be explored 
include the effects of a periodic pinning array 
at and away from commensurability, as well as
the effects of disorder on the coupling of particles confined
in adjacent two-dimensional planes.

\acknowledgments
This work was carried out under the auspices of the NNSA of the U.S. DoE
at LANL under Contract No. DE-AC52-06NA25396.

\end{document}